\documentclass[
reprint,
superscriptaddress,
amsmath,amssymb,
aps,
pra,
]{revtex4-1}
\usepackage{float}
\usepackage{xcolor}

\usepackage{graphicx}
\usepackage{dcolumn}
\usepackage{bm}
\usepackage[colorlinks=true, pdfstartview=FitV, linkcolor=blue, citecolor=blue, urlcolor=blue]{hyperref}
\usepackage{dsfont}

\usepackage[separate-uncertainty=true]{siunitx}
\usepackage[siunitx]{circuitikz}
\usepackage{tabularx}

\graphicspath{{figures/}}

\begin{document}

\preprint{APS/123-QED}

\title{Ultra-High Q Nanomechanical Resonators for Force Sensing}

\author{Alexander Eichler}
\affiliation{Laboratory for Solid State Physics, ETH Z\"{u}rich, CH-8093 Z\"urich, Switzerland.}

\date{\today}

\begin{abstract}
Nanomechanical resonators with ultra-high quality factors have become a central element in fundamental research, enabling measurements below the standard quantum limit and the preparation of long-lived quantum states. Here, I propose that such resonators will allow the detection of electron and nuclear spins with high spatial resolution, paving the way to future nanoscale magnetic resonance imaging instruments. The article lists the challenges that must be overcome before this vision can become reality, and indicates potential solutions.
\end{abstract}

	\maketitle


\clearpage
\section{Introduction}
The detection of electron and nuclear spins is an important open frontier in scanning force microscopy. The demonstration of single electron spin detection in 2004 already hinted at the capabilities of nanomechanical resonators as force sensors~\cite{Rugar_2004}. The measurement apparatus used in that experiment, however, required an integration time of 13 hours per point, which is unpractical for applications.

One goal of the community is to improve scanning force microscopy to a point where electron spin states can be detected within their coherence time or lifetime, establishing strong spin-mechanics coupling~\cite{Rabl_2009} and enabling scanning instruments to address spin quantum memories~\cite{rabl2010quantum}. A second long-term vision is force-detected nanoscale magnetic resonance imaging (NanoMRI). In this application, nuclear spins are detected over distances of tens of nanometers to form a three-dimensional image of complex molecules and other nanoscale objects~\cite{Degen_2009,Nichol_2013,Grob_2019}. Once the spatial resolution approaches the atomic scale, NanoMRI can become an important tool for structural biology.

Strong spin-mechanics coupling and NanoMRI both require nanomechanical sensors with the capability to detect zeptonewton forces within seconds, translating into a force sensitivity below \SI{1}{\atto\newton\per\hertz}. Such a sensitivity is difficult to achieve. The main obstacle to overcome is usually the thermomechanical force noise, which has a single-sided power spectral density (PSD) of $S_\mathrm{th} = 4 k_\mathrm{B} T \gamma$, where $k_\mathrm{B}$ is Boltzmann's constant, $T$ is the sensor temperature, and $\gamma = m \omega_0/Q$ is the damping coefficient defined by the effective resonator mass $m$, the resonance frequency $f_0 = \omega_0/2\pi$, and the quality factor $Q$. The spin signal to be detected is in direct competition with the displacement fluctuations $S_\mathrm{x}$ generated by this force noise, which has a frequency-dependent PSD of $S_\mathrm{x}(\omega/2\pi)=S_\mathrm{th}\chi^2(\omega/2\pi)$ with the resonator susceptibility
\begin{align}
    \chi^2(\omega/2\pi) = \frac{1}{m^2}\times \frac{1}{(\omega^2-\omega_0^2)^2 + \omega_0^2\omega^2/Q^2}\,.
\end{align}

The most widely used strategy to reduce the force noise (at a given temperature) is to reduce the resonator mass and resonance frequency, resulting in long and thin geometries. Following this approach, researchers explored top-down fabricated pendulum-style cantilevers~\cite{Mamin_2001, Tao_2014,Heritier_2018} as well as bottom-up-grown devices such as nanowires~\cite{Nichol_2012, Rossi_2017, Delepinay_2017, sahafi_2019ultralow}, carbon nanotubes~\cite{Moser_2013, deBonis_2018ultrasensitive}, and graphene sheets~\cite{weber2016force}. All of these resonators have spring constants on the order of $k = m\omega_0^2 \approx \SI{1}{\milli\newton\per\meter}$ or below, causing them to vibrate with sizeable amplitudes in response to tiny forces. At the same time, their low masses and frequencies render them susceptible to interactions that can degrade their performances. Most notably, non-contact friction sets in when the sensor tip approaches a surface, increasing the force noise PSD via a reduction of $Q$~\cite{stipe_noncontact_2001, volokitin_noncontact_2003, zurita-sanchez_friction_2004, kuehn_dielectric_2006, volokitin_near-field_2007, yazdanian_dielectric_2008, kisiel_suppression_2011, she_noncontact_2012, den_haan_spin-mediated_2015, de_voogd_dissipation_2017, Heritier_2021}. In addition, conservative tip-surface interactions can cause bending and mechanical instability of cantilevers and impede high-resolution scans~\cite{Krass_2022}. In general, the impact of non-contact friction is thought to scale roughly with $1/\omega_0$~\cite{yazdanian_dielectric_2008}, while static bending decreases with increasing spring constant as $1/k = 1/m\omega_0^2$.

\section{Dissipation Dilution}
Between 1994 and 2011, various studies investigated the impact of tensile stress on the quality factor $Q$ in doubly-clamped resonators~\cite{gonzalez1994brownian,Southworth_2009,Unterreithmeier_2010,Schmid_2011}. It was found that tensile stress can increase the angular frequency $\omega_0$ without affecting the overall damping coefficient $\gamma$ (or the mass $m$), implying that $Q \propto \omega_0$. This effect became known under the term `dissipation dilution', illustrating the fact that the dissipation $\gamma$ remains the same while the ratio $\gamma/\omega_0\propto Q^{-1}$ is reduced~\cite{Fedorov_2019}. For a force sensor, this provides an opportunity for engineering higher spring constants without any loss in sensitivity.

Dissipation dilution has become a key feature of optomechanical resonators. Quality factors in excess of $10^6$ were implemented in silicon nitride membranes~\cite{Zwickl_2008,Thompson_2008} and strings~\cite{Verbridge_2008} in 2008. Paired with optomechanical detection schemes for precise displacement measurements~\cite{Jayich_2008,Anetsberger_2009,Unterreithmeier_2009,Gavartin_2012}, these devices enabled high-impact results in fundamental~\cite{Faust_2013,Purdy_2013,Yuan_2015,Wilson_2010} and applied science~\cite{Andrews_2014,Bagci_2014}. In an additional step, phononic crystal engineering allowed researchers to reach the material dissipation limits of nanomechanical resonators by elimination of radiation to the substrate~\cite{Wilson_2009,Schmid_2011,Suhel_2012,Purdy_2012,Villanueva_2014,Chakram_2014,Tsaturyan_2014,yu_2014,Ghadimi_2017} and clamping losses~\cite{Tsaturyan_2017,Ghadimi_2018}. Further improvement was made by the realization that the dissipation dilution of a resonator can be increased by appropriate mode-shape engineering, resulting in various geometrical solutions~\cite{Norte_2016,Reinhardt_2016,Reetz_2019,Catalini_2020,Fedorov_2020,Bereyhi_2022,Bereyhi_2022_NC,Shin_2022}. In parallel, experimental demonstrations of the properties of silicon nitride resonators in extreme parameter regimes laid the basis for future applications~\cite{Rossi_2018,Mason_2019,Rossi_2019,Guo_2019,Fischer_2019,Halg_2021,Seis_2022,Halg_2022,Gieseler_2012}.

\begin{table*}[t!]
\centering
\setlength{\tabcolsep}{10pt}
\begin{tabular}{lcccccl}
\hline\hline
Reference & type & $T$ (K) & $m$ (kg) & $f_0$ (MHz) & $Q$ ($\times 10^6$) & $\gamma$ (kg/s) \\
\hline
\cite{Thompson_2008}~Thompson \textit{et al.} & membrane & $293$ & $4\times 10^{-11}$ & $0.13$ & $1.1$ & $3.1\times 10^{-11}$ \\
\cite{Zwickl_2008}~Zwickl \textit{et al.} & membrane & $0.3$ & $4\times 10^{-11}$ & $0.13$ & $11$ & $3.1\times 10^{-12}$ \\
\cite{Verbridge_2008}~Verbridge \textit{et al.} & string & $293$ & - & $1.0$ & $1.3$ & - \\
\hline
\cite{Wilson_2009}~Wilson \textit{et al.} & membrane & $293$ & - & $1.0$ & $4$ & - \\
\cite{Schmid_2011}~Schmid \textit{et al.} & string & $293$ & $1\times 10^{-12}$ & $0.17$ & $6.9$ & $1.6\times 10^{-13}$ \\
\cite{Purdy_2012}~Purdy \textit{et al.} & membrane & $293$ & - & $0.78$ & $6$ & - \\
\hline
\cite{Gavartin_2012}~Gavartin \textit{et al.} & string & $293$ & $9\times 10^{-15}$ & $2.9$ & $0.48$ & $3.4\times 10^{-13}$ \\
\cite{Faust_2013}~Faust \textit{et al.} & string & $10$ & - & $7.5$ & $0.2$ & - \\
\cite{Purdy_2013}~Purdy \textit{et al.} & membrane & $4.9$ & $7\times 10^{-12}$ & $1.6$ & $3.3$ & $2.1\times 10^{-11}$ \\
\hline
\cite{Chakram_2014}~Chakram \textit{et al.} & membrane & $293$ & - & $2.0$ & $50$ & - \\
\cite{Tsaturyan_2014}~Tsaturyan \textit{et al.} & membrane & $8$ & - & $2.7$ & $5.2$ & - \\
\cite{Yuan_2015}~Yuan \textit{et al.} & membrane & $0.01$ & $1\times 10^{-10}$ & $0.24$ & $127$ & $1.2\times 10^{-12}$ \\
\hline
\cite{Reinhardt_2016}~Reinhardt \textit{et al.} & trampoline & $293$ & $4\times 10^{-12}$ & $0.04$ & $45$ & $2.3\times 10^{-14}$ \\
\cite{Norte_2016}~Norte \textit{et al.} & trampoline & $293$ & $1\times 10^{-12}$ & $0.14$ & $98$ & $9.0\times 10^{-15}$ \\
\cite{Tsaturyan_2017}~Tsaturyan \textit{et al.} & membrane & $293$ & $1.6\times 10^{-11}$ & $0.78$ & $210$ & $3.6\times 10^{-13}$ \\
\hline
\cite{Ghadimi_2018}~Ghadimi \textit{et al.} & string & $293$ & $5\times 10^{-15}$ & $1.3$ & $800$ & $5.1\times 10^{-17}$ \\
\cite{Rossi_2018}~Rossi \textit{et al.} & membrane & $11$ & $2.3\times 10^{-12}$ & $1.1$ & $1030$ & $1.6\times 10^{-14}$ \\
\cite{Guo_2019}~Guo \textit{et al.} & string & $293$ & $7.4\times 10^{-14}$ & $0.95$ & $27$ & $1.6\times 10^{-14}$ \\
\hline
\cite{Fischer_2019}~Fischer \textit{et al.} & trampoline & $293$ & $5\times 10^{-13}$ & $0.41$ & $2.4$ & $5.4\times 10^{-13}$ \\
\cite{Gisler_2022}~Gisler \textit{et al.} & string & $0.05$ & $9\times 10^{-15}$ & $1.4$ & $2300$ & $3.6\times 10^{-17}$ \\
\cite{Seis_2022}~Seis \textit{et al.} & membrane & $0.08$ & $1.5\times 10^{-11}$ & $1.5$ & $1500$ & $9.3\times 10^{-14}$ \\
\hline
\cite{Bereyhi_2022}~Bereyhi \textit{et al.} & string (perimeter) & $293$ & $1.7\times 10^{-14}$ & $0.35$ & $3600$ & $1.0\times 10^{-17}$ \\
\cite{Bereyhi_2022_NC}~Bereyhi \textit{et al.} & string (hierarchical) & $6$ & $3.8\times 10^{-14}$ & $0.11$ & $1100$ & $2.3\times 10^{-17}$ \\
\cite{Shin_2022}~Shin \textit{et al.} & string (spiderweb) & $293$ & $5.3\times 10^{-13}$ & $0.13$ & $1800$ & $2.5\times 10^{-16}$ \\
\hline
\cite{Beccari_2022}~Beccari \textit{et al.}$^*$ & sSi string & $7$ & $8.8\times 10^{-15}$ & $1.4$ & $13000$ & $6.0\times 10^{-18}$ \\

\hline\hline
\end{tabular}
\caption{Selection of parameter values of silicon nitride devices extracted from literature and sorted by year of publication. The work by Beccari \textit{et al.} is marked with a star because it is achieved with a crystalline material (strained silicon). Note the large increase of $Q$ with time while other parameters show no systematic change.}
\label{fig:table1}
\end{table*}

Table~\ref{fig:table1} contains a selection of silicon nitride device parameters sorted by the publication year. It is worth noting that $Q$ increased from `simple' dissipation-diluted devices to the latest generation of resonators by more than three orders of magnitudes, while the typical resonance frequency and the mass remained in a similar range (for similar resonator types). This signifies a reduction of $\gamma$ and $S_\mathrm{th}$, as desired for force sensing applications. Silicon nitride resonators therefore offer the prospect of unprecedented force experiments: the best device by Bereyhi \textit{et al.}~\cite{Bereyhi_2022} has a room-temperature force noise PSD of $S_\mathrm{th} = \SI{0.16}{\atto\newton\squared\per\hertz}$. In principle, this would be enough to detect a single proton spin with a magnetic moment of $\mu_\mathrm{p} = \SI{1.4e-26}{\joule\per\tesla}$ inside a nanomagnetic field gradient of $G = \SI{6e6}{\tesla\per\meter}$ (achievable with a sharp magnetic tip~\cite{Kosata_2020}) within an averaging time of $S_\mathrm{th}/(\mu_\mathrm{p}G)^2 = \SI{22}{\second}$. The device by Gisler \textit{et al.}~\cite{Gisler_2022}, characterized in a dilution refrigerator, offers a nominal PSD of $\SI{100}{\zepto\newton\squared\per\hertz}$, reducing the required averaging time for single-proton spin detection to \SI{14}{\milli\second}.

\section{Steps to Applications}
The bare sensitivity values shown in Table~\ref{fig:table1} are impressive, but translating these numbers into real scanning force applications requires a number of steps that we address in the following.

\paragraph{Additional noise sources:} while thermal fluctuations (due to $S_\mathrm{th}$) usually dominate the noise budget of a nanomechanical sensor, there can be other important noise sources in a force sensing experiment. Most importantly, detector noise with a PSD denoted by $S_\mathrm{det}$ adds uncertainty to a measurement of the resonator displacement. Typically, $S_\mathrm{det}$ is white to a good approximation and appears in the same way as an additional force noise of the form $S_\mathrm{det}/\chi^2$.

In order to reduce $S_\mathrm{det}$, an efficient detection mechanism is required. In cavity optomechanics~\cite{Aspelmeyer_2014}, efficient readout is achieved by coupling the resonator displacement dispersively to an electromagnetic cavity. Both optical and microwave cavities have been applied to silicon nitride membranes~\cite{Jayich_2008,Wilson_2009,Yuan_2015,Andrews_2014,Bagci_2014} and strings~\cite{Anetsberger_2009,Gavartin_2012,Faust_2012}, and it is currently not clear which cavity variant will be most useful for scanning force microscopy: while optical cavities are versatile and generally offer the lowest detection uncertainty, the absorption of photons by the resonator may lead to heating at cryogenic temperatures~\cite{Gisler_2022}. Superconducting microwave cavities, on the other hand, have been tested to very low temperatures~\cite{Yuan_2015,Seis_2022}, but have the drawback that they require metallization of the resonator and cannot operate at ambient temperatures or in high magnetic fields.

As the detection uncertainty becomes low, quantum backaction sets in. The force noise PSD added by quantum backaction is given by $S_\mathrm{qba} = \hbar^2/(4 S_\mathrm{det}\eta)$~\cite{Aspelmeyer_2014}, where $\hbar$ is the reduced Planck constant and $\eta$ quantifies the measurement efficiency, which can approach unity~\cite{Rossi_2018}. Note that $S_\mathrm{qba}$ corresponds to an (approximately) white force noise, while the contribution of $S_\mathrm{det}/\chi^2$ increases with the detuning from resonance roughly as $(\omega^2 - \omega_0^2)^{2}$. For this reason, the optimal balance between the detector noise and quantum backaction depends on the bandwidth (i.e., the temporal resolution) desired in a measurement.

\begin{figure*}[t]
    \includegraphics[width=\textwidth]{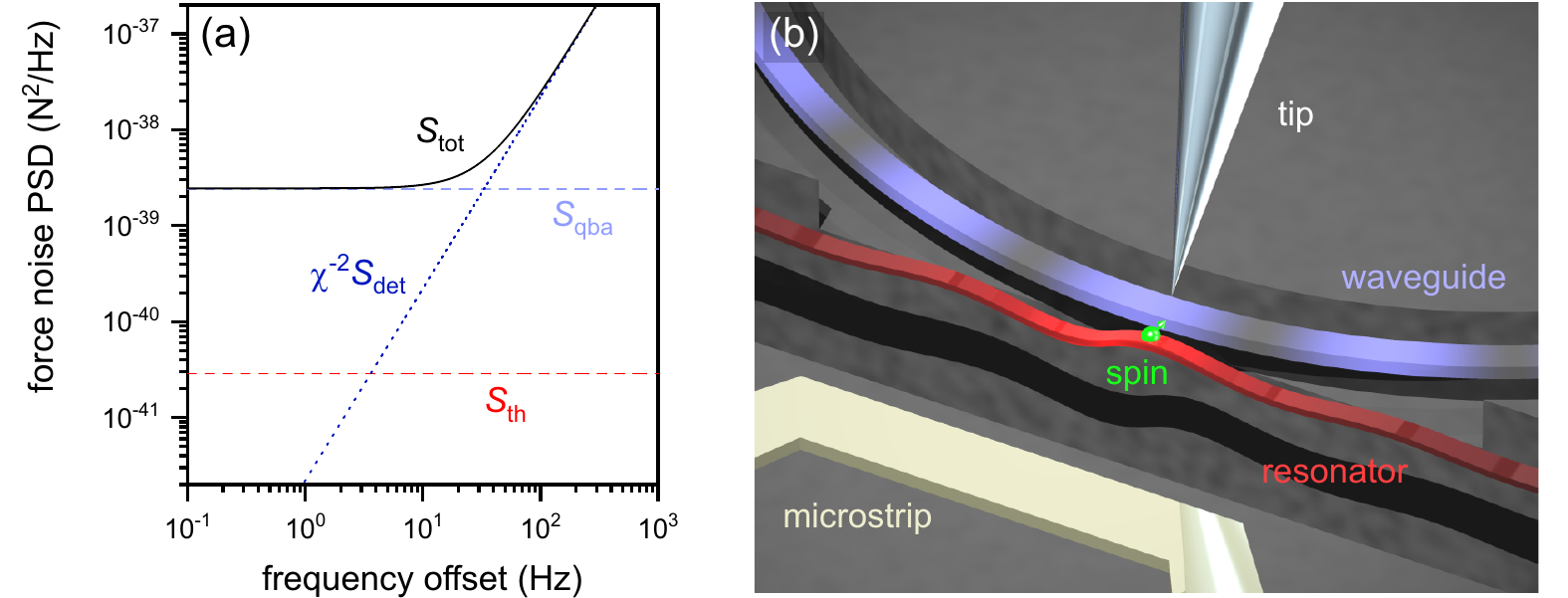}
    \caption{ (a)~Calculated force noise budget as a function of the frequency offset $\vert \omega-\omega_0\vert/2\pi$ for the perimeter device in Ref.~\cite{Bereyhi_2022}, assuming the device is passively cooled to \SI{50}{\milli\kelvin} and measured with a detection noise PSD of \SI{1e-29}{\meter\squared\per\hertz} and a measurement efficiency of $\eta = 0.5$. The sum of all contributions is shown as a solid black line labelled $S_\mathrm{tot}$. Within a bandwidth of \SI{50}{\hertz}, the total noise has a standard deviation of \SI{0.65}{\atto\newton}. (b)~Illustration of possible scanning force microscopy setup using a string resonator as sensor. A single spin (or a sample comprising a spin ensemble) is approached by a sharp, rigid scanning tip. The resulting changes in the resonator vibrations are probed with an on-chip optical waveguide whose light field is evanescently coupled to the resonator displacement. A metal microstrip is used to apply ac magnetic fields for spin inversion.}
    \label{fig:string_device}
\end{figure*}

A calculated example for different force noise contributions is shown in Fig.~\ref{fig:string_device}(a). We note that with the parameters chosen for this calculation, the total force noise is dominated by $S_\mathrm{qba}$ up to a frequency offset of about \SI{10}{\hertz}. The optimal compromise between high detection precision and low quantum backaction is termed the `standard quantum limit' (SQL). While representing a general property of interferometers, the SQL is not a fundamental limitation to quantum measurements. Various methods have been proposed and tested to enable displacement detection with an imprecision below the SQL~\cite{Braginsky_1980,Braginsky_1996,Kimble_2001,Giovannetti_2004,Clerk_2008,Lecocq_2015,Kampel_2017,Shomroni_2019,Mason_2019}. In the context of scanning force microscopy, these methods offer a strategy for improving the sensitivity beyond backaction-dominated situations such as shown in Fig.~\ref{fig:string_device}(a).

For very low temperatures, the thermal noise becomes negligible and the resonator approaches its ground-state motion set by quantum fluctuations. For the purpose of this article, quantum fluctuations are no important limit, as we expect to remain in the regime $\frac{k_\mathrm{B}T}{\hbar\omega_0}\approx n \gg 1$, where $n$ is the number of average phonons occupying the bath at the resonance frequency. Quantum fluctuations pose also no fundamental limit for feedback damping~\cite{Courty_2001}, which is often used to match the sensor bandwidth to that of the signal~\cite{Poggio_2007_PRL,Degen_2009}.

One additional aspect we need to keep in mind is non-contact friction. Even at MHz frequencies, this effect can eventually become dominant when all other noise sources are sufficiently small. Previous experiments found an added non-contact dissipation coefficient of $\gamma_\mathrm{ncf} \approx \SI{5e-14}{\kilo\gram\per\second}$ at room temperature and at tip-surface distances around \SI{1}{\nano\meter}~\cite{Halg_2021,Halg_2022}. It is to be expected that this value can be significantly reduced at cryogenic temperatures, larger tip-surface distances, and with suitable surface treatment techniques that minimize the presence of adsorbant layers~\cite{Heritier_2021}.

\paragraph{Instrumentation:} to become a scanning force sensor, the resonator must be mounted in a dedicated measurement apparatus. In contrast to singly-clamped cantilevers, doubly-clamped beams or membranes are more difficult to utilize as scanning probes. This problem can be solved by inverting the typical geometry found in most atomic force microscopes: using the resonator as a sample plate and as the vibrating sensor at the same time, a sharp (but rigid) scanning tip is employed to probe interactions with the samples~\cite{Halg_2021,Halg_2022}. The same design can potentially be used with trampolines and strings, as long as the resonator surface is large enough to allow sample deposition, see Fig.~\ref{fig:string_device}(b). In the case of NanoMRI, the relevant sample scale is on the order of \SI{100}{\nano\meter}~\cite{Degen_2009,Krass_2022}.

\paragraph{Sample deposition:} suitable methods to load samples onto fragile resonators must be developed. One possibility is the direct deposition from liquid droplets through micropipettes~\cite{Halg_2021} or hollow cantilevers~\cite{Meister_2009}. The main drawback of this method is the creation of residues from the liquid that can potentially reduce the resonator's sensitivity. Alternative methods that avoid contamination include laser-induced desorption of particles from a substrate in vacuum~\cite{Bykov_2019} or spray injection through a finely tuned valve~\cite{Chaste_2012,Tavernarakis_2014}. For biological samples, freeze-drying is advantageous to avoid structural damage during the chamber evacuation~\cite{Giddings_1986}. However, it appears unlikely that nanomechanical resonators can survive direct dipping into liquid nitrogen. This issue has previously been overcome by a modular sample preparation procedure for cantilevers~\cite{Overweg_2015,Krass_2022}.

\paragraph{Sensing protocols:} in this last section, we take a look at some spin detection protocols. In general, moving from topography scanning force microscopy to the special case of spin detection requires two additional components. The first of these is a magnetic field gradient $G$ that translates a magnetic moment $\mu_\mathrm{p}$ into a measurable force $\mu_\mathrm{p}G$. The best nanomagnets used in previous NanoMRI demonstrations yielded values between $G = \SI{2.3e6}{\mega\tesla\per\meter}$~\cite{Grob_2019} and \SI{6e6}{\mega\tesla\per\meter}~\cite{Mamin_2012} at distances of several tens of nanometers. Nanoscale magnets used in harddisk write heads even reach \SI{28e6}{\mega\tesla\per\meter}~\cite{Tao_2016}, indicating the potential for further improvements in NanoMRI. From simulations, the field generated by a magnetic coating layer on a sharp tip, as shown in the illustration in Fig.~\ref{fig:string_device}(b), is expected to reach \SI{6e6}{\mega\tesla\per\meter} or more~\cite{Kosata_2020}.

The second component that is indispensable in most NanoMRI protocols is a source for oscillating magnetic field pulses. These pulses are used to periodically manipulate the spins at their Larmor frequencies (in the MHz to GHz range) to induce spin-mechanics coupling. In order to minimize the amount of electric current needed to create the magnetic field pulses, and hence the heating imposed on the instrument, it is advantageous to place the field source as close as possible to the sample. Metallic microstrips have become a popular choice because they can easily be fabricated on chips and cover a wide range of frequencies~\cite{Poggio_2007}.

With cantilevers in the kHz range, various protocols have been developed and tested~\cite{Sidles_1991,Mamin_2003,Rugar_2004,Mamin_2007,Degen_2009,Poggio_2010,Wagenaar_2017}. They typically rely on periodic spin inversions (induced by ac magnetic field pulses) to create a force acting on the cantilever. When contemplating resonators in the range of \SI{100}{\kilo\hertz} to \SI{10}{\mega\hertz} as force sensors, most of these protocols are impractical because adiabatic inversion of nuclear spins at such high frequencies requires too much power to be implemented in a cryogenic environment.

A technique developed especially for nanowires at a frequency close to \SI{1}{\mega\hertz} makes use of switchable magnetic field gradients to create a force resonant with the sensor~\cite{Nichol_2012,Nichol_2013}. This protocol could be translated to membrane and string resonators if switchable magnetic tips can be fabricated. Alternatively, it was proposed to invert spins at the \textit{frequency difference} between two modes, giving rise to parametric coupling~\cite{Dougherty_1996,Moore_2010}. Membrane resonators can be engineered to have two normal modes split by a few kHz in frequency~\cite{Catalini_2020,Halg_2022}, which offers a convenient tool for nuclear spin detection~\cite{Kosata_2020}. Finally, nanoscale resonant coupling between nuclear spins and a mechanical resonator, as originally envisioned~\cite{Sidles_1992_PRL}, could at long last be realized with this new class of sensors. Resonant coupling allows not only to detect nuclear spins, but also to manipulate them via the mechanics. Such manipulations have already been demonstrated with the electron spin ensembles of large atomic clouds~\cite{Karg_2020,Thomas_2021}.

In general, ensembles of electron spins are more accessible to mechanical detection than nuclear spins, both due to their larger magnetic moment and because adiabatic inversion can be achieved much faster. For this reason, pioneering experimental demonstrations of spin detection with membrane~\cite{Scozzaro_2016} and trampoline~\cite{Fischer_2019} resonators targeted electron spins in diphenyl picrahydrazyl samples with a size of a few tens of micrometers. These experiments, albeit not on the nanoscale, fuel our optimism that silicon nitride resonators will make excellent NanoMRI sensors.

\section{Outlook}
Looking forward, we can only speculate what the next step in the evolution of nanomechanical sensors will be. Pushing the limits to lower temperatures or cleaner surfaces will certainly allow for yet another increase in sensitivity. A major breakthrough could be achieved if the microscopic origin of intrinsic nanomechanical dissipation is precisely understood and can be addressed. With regards to concrete spin detection protocols, the possibilities afforded by geometrical design and pulsed spin control are far from exhausted. Here, we will hopefully see a host of interesting new ideas over the next few years.

Much remains to be done before silicon nitride membranes or strings can be used for scanning force microscopy, and in particular for nuclear spin detection and NanoMRI. Nevertheless, the prospects offered by this new class of mechanical sensors are truly encouraging. With all of the exciting developments taking place in the optomechanics community, it may well be that this new material paves the way to realizing an old dream.

\acknowledgements The author would like to thank Nils Engelsen and Letizia Catalini for critical reading and feedback on the manuscript.

\providecommand{\noopsort}[1]{}\providecommand{\singleletter}[1]{#1}%
\end{document}